# Vehicular Ad-hoc Networks: Architecture, Applications and Challenges

**Taoufik Yeferny and Sofian Hamad**

College of Science, Northern Border University, Arar, Saudi Arabia

**Summary**
With the emergence of Information and Communication Technologies (ICT) and wireless embedded sensing devices into modern vehicles, Intelligent Transport System (ITS) becomes a reality and an indispensable component of smart cities. The purpose of ITS is to improve road safety and traffic efficiency as well as offering infotainments services. In fact, warning drivers in the right time about dangerous situations on the road and providing them with prior information about traffic will undoubtedly leads to enhance driver's safety and reduce traffic congestion. Technically speaking, ITS is based on self-organizing wireless networks, known as vehicular ad-hoc networks (VANETs). Mobile vehicles in VANET might play the role of stationary sensors in infrastructure-based networks. They can detect, gather and disseminate real-time data about traffic, driving conditions and potential hazards on roads. In this respect, we review in this study, recent developments on the design of VANET protocols and applications. We first introduce the architecture of VANETs then we review their unique characteristics and applications. Thereafter, we discuss the main research challenges and open issues to be considered for designing efficient and a cost-effective VANET protocols and applications.

## 1. Introduction

Recently, with the emergence of Information and Communication Technologies (ICT) and wireless embedded sensing devices into modern vehicles, Intelligent Transport System (ITS) becomes a reality and an indispensable component of smart cities. The purpose of ITS is to improve road safety and traffic efficiency as well as offering infotainments services. In fact, warning drivers in the right time about dangerous situations on the road and providing them with prior information about traffic will undoubtedly leads to enhance driver's safety and reduce traffic congestion. Technically speaking, ITS is based on self-organizing wireless networks [1], [2], [3], [4], known as vehicular ad-hoc networks (VANETs) [5]. In these mobile networks, vehicles are self-organized and communicate each with other without requiring any infrastructure or centralized coordinating entity [6]. Mobile vehicles in VANET might play the role of stationary sensors in infrastructure-based networks. They can detect, gather and disseminate real-time data about traffic, driving conditions and potential hazards on roads. In fact, disseminating such data in VANET would enhance road safety and driving comfort [7], [8]. Due to the characteristics of VANETs, such as the high mobility of vehicles, network partitions and fragmentation, information must be exchanged in an efficient way to avoid the well-known broadcast storm problem [9], [10], [11]. In this respect, designing efficient protocols and application for VANET has become a core and challenging issue, which has attracted the interest of researchers, networking professionals and automotive companies [12], [13], [14], [15]. In this study, we review recent developments on the design of VANET protocols and applications. We first introduce the architecture of VANETs then we review their unique characteristics and applications. Thereafter, we discuss the main research challenges and open issues to be considered for designing efficient and a cost-effective VANET protocols and applications.

The remainder of this paper is structured as follows. In Section 2, we introduce the architecture of VANET, including its main components and interactions between them. The unique VANET characteristics are discussed in Section 3 followed by detailed description of VANET applications in Section 4. In Section 5, we discuss the main research challenges and open issues to be considered for designing efficient and a cost-effective VANET protocols and applications. Finally, Section 6 concludes this paper and pins down some research directions and perspectives.

## 2. VANET Architecture

In what follows, we thoroughly describe the main VANET components and the interactions between them.

### 2.1 Main components

With respect to the IEEE 1471-2000 [16] and ISO/IEC 42010 [17] standards, VANET components are classified into three domains:

- **Mobile domain** includes the vehicle and the mobile device domains. The former comprises all type of vehicles (e.g., cars, trains, buses). The latter includes all types of portable devices (e.g., smartphones, laptop, smart watches).
- **Infrastructure domain** incorporates the roadside infrastructure domain (e.g., traffic light, camera, etc.) and the central infrastructure domain (e.g., Traffic Management Centres (TMCs), Vehicle Management Centres).
- **Generic domain** includes the Internet and the Private infrastructures.

The European architecture standard for VANETs is little different. In fact, it relies on the CAR-2-X communication system pursued by the CAR-2-CAR Communication Consortium [18]. As shown in Figure 1, the reference architecture of the C2C Communication System, comprises the following domains:

- **In-vehicle domain** is composed of one or multiple application units (AUs) and one On-Board Unit (OBU). An AU is a dedicated device, which can be an integrated part of a vehicle or a separate portable device such as smartphone, laptop, etc. It runs one or many applications that exploit the OBU communication capabilities. The AUs and OBU are permanently connected through a wired or wireless connection.
- **Ad-hoc domain** is composed of vehicles equipped with OBUs and stationary Road-Side Units (RSUs) deployed in specific locations along the road. OBUs can communicate each with other, directly or via multi-hop, using wireless short-range communication devices allowing ad-hoc communications between vehicles. An RSU is a stationary device that can be connected to an infrastructure network or to the Internet. It can send, receive or forward data in the ad-hoc domain (i.e., vehicles equipped with OBUs and RSUs), which enables to extend the coverage of the ad-hoc network. An OBU may access to the Internet via an infrastructure connected RSU, public commercial or private wireless Hot Spots (HSs) to communicate with Internet nodes or servers.
- **Infrastructure domain** access consists of HSs and RSUs. In case that neither RSUs nor HSs provide Internet access, OBUs can exploit cellular radio networks for example HSDPA, WiMax and 4G.

2.2. Communication Architecture

VANET communications are categorized as follows [7]:

- **In-vehicle communication** between OBU of the vehicle and its AUs.
- **Vehicle-to-vehicle (V2V)** wireless communications between vehicles via their OBUs.
- **Vehicle-to-infrastructure (V2I)** refers to bidirectional wireless communications between vehicles and infrastructure-connected RSUs.
- **Infrastructure-to-Infrastructure (I2I)** communications between RSUs enable extending the coverage of the network.
- **Vehicle-to-broadband cloud (V2B)** communications between vehicles and broadband cloud via wireless broadband technologies such as 3G/4G.

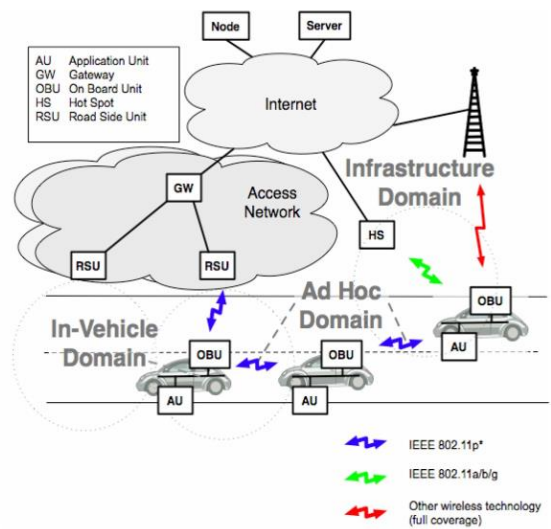

Fig. 1 C2C-CC reference architecture

## 2. Characteristics of VANETs

VANETs are special type of ad-hoc networks distinguished by the following characteristics derived from [19] and [20]:

- **Mobility**: VANETs are mainly composed of fixed RSUs and moving vehicles. Vehicle's speed varies from very low to very high, leading to new communication challenges. Indeed, in areas of high traffic jam, vehicles are stopped or moving slowly and therefore they have enough time to exchange messages. However, they face major challenges due to the high density of vehicles such as data collision, channel fading, message dropping and other interference problems. In areas of low traffic (e.g., highway), vehicle speed is very high leading to others communication challenges such as small communication window (few seconds), link failures, high end-to-end

(ETE) delay, etc.
- **Movement pattern**: node movement in VANETs differs from that of Mobile Ad-hoc Networks (MANETs). In fact, in MANETs, mobile nodes are free to move anywhere at any time. However, in VANET, vehicles follow the topology of road networks of the geographic areas where they drive. In general, there are three situations: urban area, rural area and highway. As shown in Figure 2, the urban area has more complex road network, denser in terms of vehicles number than the rural area. Furthermore, it contains more obstacles, traffic signals and RSUs whenever compared to rural area and highway. In the latter, vehicles move in one direction over many lanes. The spatial attributes of the road network have an impact on the communication efficiency and effectiveness.

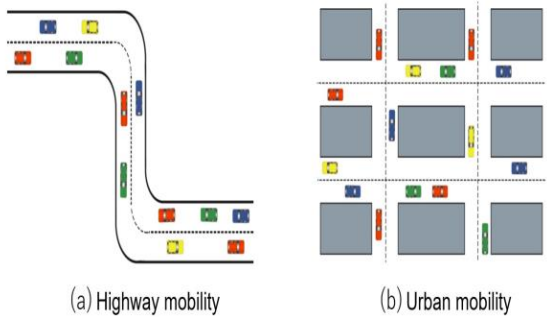

Fig. 2  Vehicle mobility

- **Traffic density**: it ranges from high to low density, depending on the geographic area (i.e., high traffic density in urban area and low traffic density in rural area and highway) and the time factor (i.e., low traffic density during off-peak hours and high traffic during rush hours). Traffic density raises crucial challenges related to the design of efficient VANET communication protocols. For instance, in rural areas with very low traffic density, data dissemination protocols must deal with the network disconnection issue. However, advanced data dissemination mechanisms should be used to avoid the well-known broadcast storm issue in the case of very high traffic density, especially in urban area during rush hours.
- **Heterogeneity**: VANET nodes have different characteristics and capabilities. For instance, vehicles are moving nodes, which have different communication ranges, sensing capabilities and categories (i.e., private, authority and maintenance vehicles). Whereas RSUs are stationary nodes placed in some pertinent locations and equipped with complete ad-hoc features.

## 3. VANET Applications

According to [19], VANET applications are categorized as safety, traffic efficiency and infotainment applications. Figure 3 illustrates a taxonomy of VANET applications and related use cases.

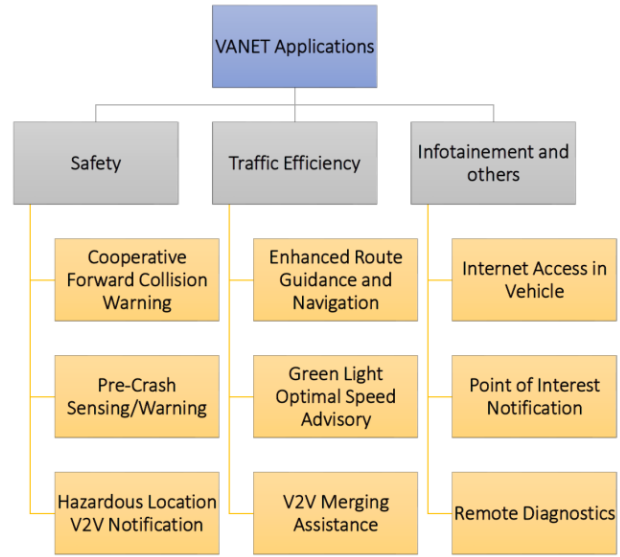

Fig. 3  Taxonomy of VANET applications

The following provides a description of the main application use cases in each category.

### 3.1 Safety Applications

Safety applications aim to warn drivers at the right time about dangerous situations in the road in order to enhance driving safety. In [18], some examples of safety use cases and their related requirements are introduced. In the following, we provide a brief description of the main use cases:

- **Cooperative Forward Collision Warning:** the goal of this use case is to avoid rear-end collisions with other vehicles by providing assistance to drivers. In fact, rear-end collisions are generally caused by driver disturbance or sudden braking. To avoid a crash, concerned vehicles share relevant information such as position, speed and direction. When a critical situation (e.g., insufficient safety distance) is detected, the vehicle warns its driver.

- **Pre-Crash Sensing/Warning:** Unlike the Cooperative Forward Collision Warning, this use case assumes that a crash is unavoidable and will take place [22]. In such cases, the involved vehicles exchange in an efficient way the related information with neighbouring vehicles in order to enable a better usage of vehicle's actuators.

- **Hazardous Location V2V Notification:** the goal of this use case is to share pertinent information about dangerous locations on the roadway (e.g., potholes, bottleneck, etc.) between vehicles in certain area. To this end, the vehicle detecting a dangerous location uses the information for optimizing its safety systems then broadcasts it to neighbouring vehicles in the surrounding area. Through V2V communications, the information is then progressively shared with other concerned vehicles. Information about dangerous locations on the roadway can also be transmitted from external service providers to RSUs, which in turn send it to some vehicles in their communication ranges. Thereafter, vehicles receiving the information can disseminate it to others via V2V communications.

### 3.2. Traffic efficiency applications

Traffic efficiency applications aim to enhance the efficiency of transportation systems by providing traffic related information to drivers or road operators. In order to achieve this goal, traffic information should be exchanged through the VANET. Therefore, road-users and road operators will respectively benefit from shorter travel times as well as from reduced costs of roads construction and maintenance. In the following, we provide a brief description of some traffic efficiency use cases introduced in the CAR-2-CAR Communication Consortium [18].

- **Enhanced Route Guidance and Navigation**: it enables the infrastructure owner to collect traffic data of a large region to be used later for predicting traffic congestion on roadways. Predicted information will be then transmitted to vehicles via RSUs. Hence, driver will be notified about the current and the expected traffic throughout the region, expected delays to reach his destination and better routes that might exist to avoid some congested roads. This will undoubtedly lead to improve the overall efficiency of the transportation system.

- **Green Light Optimal Speed Advisory**: it provides information related to the location of a signalized intersection and the signal timing (i.e., time to switch the light signal) to vehicles approaching the intersection, which contributes to smoother driving and avoid stopping. Receiving such information at the right time, the vehicle can calculate the optimal speed to reach the intersection when the traffic signal is green, and therefore the driver will not have to decrease the speed of the vehicle or to stop. This fact will probably bring about a significant increase in the traffic flow and fuel economy.

### 3.3. Infotainment and others

Non safety or traffic efficiency use cases are classified in this category. Some of these use cases provide entertainment or information on a regular basis to drivers. Other ones are transparent to the driver and they play important role for improving the vehicle functions.

- **Internet Access in Vehicle**: it allows drivers and eventually passengers to access the Internet via the VANET. In this case, RSUs act as internet gateways.

- **Point of Interest Notification**: it allows traders and advertisement companies to advertise their business promotions to nearby vehicles. To this end, an RSU broadcasts the advertisement information (e.g., location, hours of operation and pricing) to the contacted vehicles. The received advertisements will be filtered by each vehicle with respect to the driver profile and context then appropriate advertisements are presented to the driver.

## 4. Open Issues and Challenges

This section addresses the main research challenges and open issues to be considered for designing efficient and a cost-effective protocols and applications for VANET.

### 4.1 Data Dissemination in VANETS

Data dissemination is a key and hot topic in vehicular networks. The challenge is to inform vehicles about interesting events while avoiding the broadcast storm problem. In figure 4, we provide a new taxonomy of the existing protocols, which we classify as: broadcast or geocast protocols. The main goal of broadcast protocols is to deliver a given event message to all vehicles in the network without exception. However, the aim of geo-cast protocols is to disseminate the event message to a target set of vehicles driving in a geographical area, named Zone of Relevance (ZOR) [21], [22]. In the following, we introduce crucial challenges facing the design efficient data dissemination protocols.

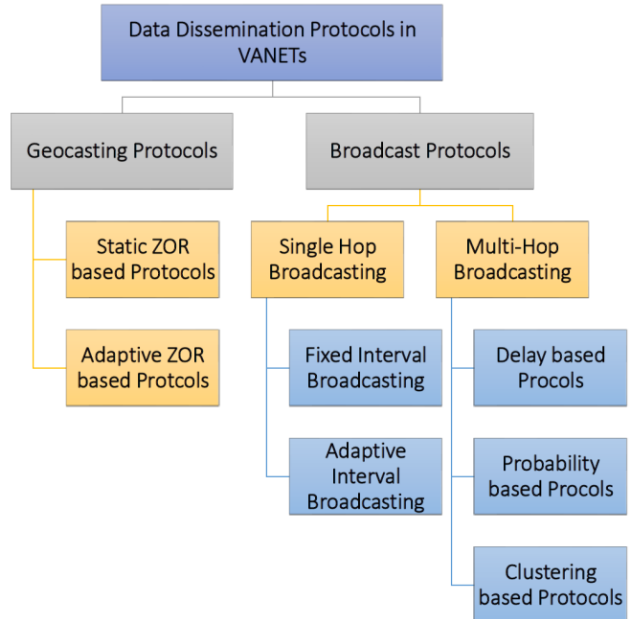

Fig. 4. Taxonomy of Data Dissemination Protocols in VANETs

- **Zone of Relevance Computation**: Efficient data dissemination of safety events could be done through geocast approach, which requires delivering information to vehicles inside the ZOR. In fact, the most important step of any geocast-based data dissemination protocol is the computation of the ZOR for a given event. Big data collected from vehicles and RSUs hides hidden mobility patterns that could be of use to design an efficient data dissemination approach and more specifically to solve the ZOR computation problem [57].
- **Event lifetime management**: In VANET, the disseminated message should be kept alive in the ZOR for a certain delay to inform new arriving vehicles. Therefore, new mechanisms must be introduced to keep relevant events alive inside the ZOR, without the need of rebroadcasting event messages as the existing approaches do.

### 4.2. Data Aggregation in VANET

As VANET is a very dynamic network due the high mobility of vehicles, information must be exchanged between mobile vehicles in an efficient way by avoiding as far as possible the broadcast storm problem. In this respect, data aggregation appears as an interesting approach allowing to integrate several data about similar events to generate a summary (or aggregate) leading to reduce network traffic. Therefore, the design of an efficient data aggregation approach that combines correlated traffic information is a challenging issue [26], [27]. The data aggregation process must deal with the following challenges:

- How to decide if two or more messages must be aggregated or not?
- How to select timely data to be aggregated?
- How to consider data from far vehicles?
- How to filter the unnecessary and duplicated messages to avoid affecting the accuracy of the shared traffic information?
- How to take into consideration the road traffic signals and speed limitations?

### 4.3. RSUs Deployment in VANETs

In VANET, RSUs can collect real-time data, from moving vehicles, then transmit it to traffic information centres for further analysis. In addition, RSUs disseminate important information that they receive from the traffic management centres to vehicles. However, communication with enormous number of vehicles driving in a big city requires the deployment of large number of RSUs in different locations [9], [28], [29]. In this respect, an efficient RSUs deployment strategy must be introduced. The challenge is to (i) minimize the number of RSUs as much as possible in order to reduce the deployment cost; and (ii) to maximize the coverage ratio.

## 5. Conclusion

In this study, first we have presented an overview of VANETs, including a detailed description of VANETs architecture, characteristics and applications. Thereafter, we have discussed the main research challenges and open issues to be considered for designing efficient and a cost-effective VANET protocols and applications. Indeed, pertinent perspectives, challenges, shortcomings in the area and some related research directions have been introduced.

**Acknowledgments**

This project was funded by Deanship of Scientific Research, Northern Border University for their financial support under grant no. SCI-2018-3-9-F-7683. The authors, therefore, acknowledge with thanks DSR technical and financial support.

**References**
[1] T. Yeferny, K. Arour, and A. Bouzeghoub, "An efficient peer-to-peer semantic overlay network for learning query routing," in Advanced Information Networking and Applications (AINA), 2013, pp. 1025–1032.

[2] T. Yeferny and K. Arour, "Efficient routing method in p2p systems based upon training knowledge," in Advanced Information Networking and Applications Workshops (WAINA), 2012, pp. 300–305.

[3] K. Arour and T. Yeferny, "Learning model for efficient query routing in p2p information retrieval systems," Peer-to-Peer Networking and Applications, vol. 8, no. 5, pp. 741–757, 2015.

[4] T. Yeferny, K. Arour, and Y. Slimani, "Routage semantique des requetes dans les systemes pair-a-pair."` CORIA, vol. 9, pp. 131–147, 2009.

[5] M. Chaqfeh, A. Lakas, and I. Jawhar, "A survey on data dissemination in vehicular ad hoc networks," Vehicular Communications, vol. 1, no. 4, pp. 214–225, 2014.

[6] W. Liang, Z. Li, H. Zhang, S. Wang, and R. Bie, "Vehicular ad hoc networks: architectures, research issues, methodologies, challenges, and trends," International Journal of Distributed Sensor Networks, vol. 11, no. 8, p. 745303, 2015.

[7] A. Awang, K. Husain, N. Kamel, and S. Aïssa, "Routing in vehicular ad-hoc networks: A survey on single-and cross-layer design techniques, and perspectives," IEEE Access, vol. 5, pp. 9497–9517, 2017.

[8] S. Allani, T. Yeferny, R. Chbeir, and S. B. Yahia, "Dpms: A swift data dissemination protocol based on map splitting," in Computer Software and Applications Conference (COMPSAC), vol. 1, 2016, pp. 817–822.

[9] T. Yeferny and S. Allani, "MPC: A rsus deployment strategy for vanet," Int. J. Communication Systems, vol. 31, no. 12, 2018.

[10] S. Hamad and T. Yeferny, "A smart data dissemination protocol for vehicular ad-hoc networksamad," International Journal of Computer Science and Network Security, vol. 19, p. 176, 2019.

[11] M. Chahal, S. Harit, K. K. Mishra, A. K. Sangaiah, and Z. Zheng, "A survey on software-defined networking in vehicular ad hoc networks: Challenges, applications and use cases," Sustainable cities and society, vol. 35, pp. 830–840, 2017.

[12] M. S. Talib, A. Hassan, B. Hussin, and A. A.-h. Hassan, "Vehicular ad-hoc networks: Current challenges and future direction of research," Jour Adv Res. Dyn. Control Syst, vol. 10, no. 2, pp. 2065–2074, 2018.

[13] I. Ahmad, R. M. Noor, I. Ahmedy, S. A. A. Shah, I. Yaqoob, E. Ahmed, and M. Imran, "Vanet–lte based heterogeneous vehicular clustering for driving assistance and route planning applications," Computer Networks, vol. 145, pp. 128–140, 2018.

[14] S. Latif, S. Mahfooz, B. Jan, N. Ahmad, Y. Cao, and M. Asif, "A comparative study of scenario-driven multi-hop broadcast protocols for vanets," Vehicular Communications, vol. 12, pp. 88–109, 2018.

[15] S. Allani, T. Yeferny, R. Chbeir, and S. B. Yahia, "A novel vanet data dissemination approach based on geospatial data," Procedia Computer Science, vol. 98, pp. 572–577, 2016.

[16] I. A. W. Group, "Ieee std 1471-2000, recommended practice for architectural description of software-intensive systems," IEEE, Tech. Rep., 2000.

[17] ISO/IEC/(IEEE), "ISO/IEC 42010 (IEEE Std) 1471-2000: Systems and Software engineering - Recomended practice for architectural description of software-intensive systems," p. 23, 07 2007.

[18] R. Baldessari, B. Bodekker, M. Deegener, A. Festag, W. Franz, C. C.¨ Kellum, T. Kosch, A. Kovacs, M. Lenardi, C. Menig et al., "Car-2-car communication consortium-manifesto," 2007.

[19] G. Karagiannis, O. Altintas, E. Ekici, G. Heijenk, B. Jarupan, K. Lin, and T. Weil, "Vehicular networking: A survey and tutorial on requirements, architectures, challenges, standards and solutions," IEEE Communications Surveys Tutorials, vol. 13, no. 4, pp. 584–616, 2011.

[20] E. Schoch, F. Kargl, M. Weber, and T. Leinmuller, "Communication patterns in vanets," IEEE Communications Magazine, vol. 46, no. 11, pp. 119–125, 2008.

[21] R. Kumar, M. Dave et al., "A review of various vanet data dissemination protocols," International Journal of u-and e-Service, Science and Technology, vol. 5, no. 3, pp. 27–44, 2012.

[22] B. Bako and M. Weber, "Efficient Information Dissemination in VANETs," in Adv. Veh. Netw. Technol., 2011, p. 20.

[23] S. Allani, T. Yeferny, and R. Chbeir, "A scalable data dissemination protocol based on vehicles trajectories analysis," Ad Hoc Networks, vol. 71, pp. 31 – 44, 2018.

[24] P. Yang, J. Wang, Y. Zhang, Z. Tang, and S. Song, "Clustering algorithm in vanets: A survey," in 9th international conference on anticounterfeiting, security, and identification (ASID). IEEE, 2015, pp. 166–170.

[25] L. Liu, C. Chen, T. Qiu, M. Zhang, S. Li, and B. Zhou, "A data dissemination scheme based on clustering and probabilistic broadcasting in vanets," Vehicular Communications, vol. 13, pp. 78–88, 2018.

[26] S. Allani, R. Chbeir, T. Yeferny, and S. B. Yahia, "Smart directional data aggregation in vanets," in IEEE 32nd International Conference on Advanced Information Networking and Applications (AINA). IEEE, 2018, pp. 63–70.

[27] B. Pan, H. Wu, and J. Wang, "Fl-asb: A fuzzy logic based adaptive period single-hop broadcast protocol," International Journal of Distributed Sensor Networks, vol. 14, no. 5, 2018.

[28] S. B. Chaabene, T. Yeferny, and S. B. Yahia, "A roadside unit placement scheme for vehicular ad-hoc networks," in International Conference on Advanced Information Networking and Applications. Springer, 2019, pp. 619–630.

[29] Ben Chaabene, Seif, Taoufik Yeferny, and Sadok Ben Yahia. "A roadside unit deployment framework for enhancing transportation services in Maghrebian cities." Concurrency and Computation: Practice and Experience, e5611, 2019.

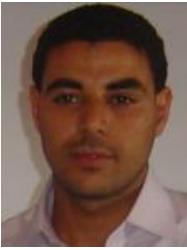
**Taoufik Yeferny** He received the M.C.S. and Ph.D. degrees in computer sciences from the University of Tunis El Manar, Tunisia, in 2009 and 2014, respectively. From 2013 to 2016, he was an Assistant Professor with the High Institute of Applied Languages and Computer Science of Beja, Tunisia. Since 2016, he has been an Assistant Professor with Northern Border University, Saudi Arabia. His current research interests include mobile P2P systems, vehicle ad hoc networks, and intelligent transportation systems.

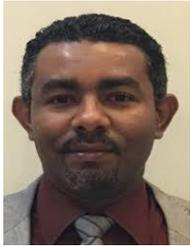
**Sofian Hamad** He received the B.Sc. degree (Hons.) in computer science from Future University, Khartoum, Sudan, in 2003, the M.Sc. degree in management business administration (MBA) from the Sudan Academy of Science, in 2007and the Ph.D. degree in electrical engineering from Brunel University, London, U.K., in 2013. He is currently an Assistant Professor with the Department of Computer Science, Northern Border University. His current research interests include ad-hoc and mesh wireless networks, vehicular technology, and the Internet of Things.